\documentclass[aps,floatfix,preprint,groupedaddress, superscriptaddress]{revtex4-1}

\usepackage{graphicx}
\usepackage{amsmath, amssymb}
\usepackage{subfigure}
\usepackage{xcolor}
\usepackage{makecell}

\providecommand{\abs}[1]{\left|#1\right|}

\begin{document}

\title{Study on Estimating Quantum Discord by Neural Network with Prior Knowledge}

\author{Yong-Lei Liu}
\email{liuyl15@mail.ustc.edu.cn}
\affiliation{Department of Modern Physics, University of Science and Technology of China, Hefei, China, 230026}

\author{An-Min Wang}
\email{Correspondence author, anmwang@ustc.edu.cn}
\affiliation{Department of Modern Physics, University of Science and Technology of China, Hefei, China, 230026}

\author{Guo-Dong Wang}
\affiliation{Department of Modern Physics, University of Science and Technology of China, Hefei, China, 230026}
\author{Yi Sun}

\affiliation{School of Computer and Control Engineering, University of Chinese Academy of Sciences, Beijing, China, 100049}
\author{Peng-Fei Zhang}
\affiliation{Department of Modern Physics, University of Science and Technology of China, Hefei, China, 230026}

\begin{abstract}
Machine learning has achieved success in many areas because of its powerful fitting ability, so we hope it can help us to solve some significant physical quantitative problems, such as quantum correlation. In this research we will use neural networks to predict the value of quantum discord. 
Quantum discord is a measure of quantum correlation which is defined as the difference between quantum mutual information and classical correlation for a bipartite system. Since the definition contains an optimization term, it makes analytically solving hard. For some special cases and small systems, such as two-qubit systems and some X-states, the explicit solutions have been calculated. However, for general cases, we still know very little. 
Therefore, we study the feasibility of estimating quantum discord by machine learning method on two-qubit systems. In order to get an interpretable and high performance model, we modify the ordinary neural network by introducing some prior knowledge which come from the analysis about quantum discord.
Our results show that prior knowledge actually improve the performance of neural network.

\keywords{quantum discord; machine learning; neural network}
\pacs{03.67.-a, 03.65.Ta}

\end{abstract}

\maketitle

\section{Introduction}
\label{intro}
In recent years, the field of machine learning (ML) has developed rapidly with deep learning, such as the famous Alpha GO\cite{agom_ts,agom_nohk} and various technologies based on computer vision.
This method also be used to solve another problem beyond computer science, such as phase transition in condensed matter\cite{nld_lptc,Yi_qlt,K_mlpscf,PB_fsp}, effective representation of quantum multi-body state\cite{gx_erqmbs}, molecular nature prediction\cite{RTML_prl2012,M_njp2013,H_jctc2013} and so on.
Thus, we hope this method will help us to deal with some quantum information problems.
In quantum information area, distinguishing a quantum state entangled or separeable is an important problem. 
Recognizing a bipartite state is entangled or separable by machine learning method\cite{belli_ml,Lu_sec,GQJ_emlqs} has begun to study.
Thus, the problem how to measure entanglement is more worthy of attention, especially for quantum correlation of mixed states. Our work attempts to predict a measure of quantum correlation.

In 2000, Olliver and Zurek\cite{qd_oz,qd_z} produced quantum discord (QD) by analogy of classical mutual information to define this measure of quantum correlation.
Because of the optimization problem in the definition of QD, it is very hard to get analytical results for general cases. So, we investigate how to apply ML technology to estimate this measure of quantum correlation. 
In ML, give a set of data $\mathcal{D}$ from a function $f:\vec{x} \mapsto y$, ML algorithm $\mathcal{A}$ can get an approximate function $h$ from a hypothesis set $\mathcal{H}$, by training based on the observed data. The function $f$ is called target function.
We hope to find an universal method to estimate the QD well, and it will work for any bipartite system. In theory, deep neural network (DNN) can fit any target function, however, extracting specific information form such model is hard due to the black box property of DNN. Thus we must modify the model to make extracting information easy by intrducing prior knowledge which base on the analysis about QD.

In Sect.~\ref{sec:qd}, we will introduce the quantum discord of two-qubit systems.
In Sect.~\ref{sec:mlm}, we will describe the neural network models we used.
In Sect.~\ref{sec:res}, we will show results in two different cases.
Finally, we discuss some problems and future works in Sect.~\ref{sec:diss}.

\section{QD of Two-Qubit}
\label{sec:qd}


QD is a measure of quantum correlation mainly for mixed states.
In a bipartite quantum system, there are classical correlation and quantum correlation. Assume $\rho^{AB}$ is the density operator of the bipartite system, and $\rho^A$, $\rho^B$ are density operators of its two part A(lice) and B(ob).
By analogy classical mutual information, Olliver and Zurek gave the definition of quantum mutual information as\cite{qd_oz}
\begin{equation}\label{eq:qmi}
  \mathcal{I}(\rho^{AB}) = S(\rho^A) + S(\rho^B) - S(\rho^{AB})
\end{equation}
where $S(\rho)=-\mathrm{tr} (\rho \log_2 \rho) = -\sum_i \lambda_i \log_2 \lambda_i$ is von Neumann entropy, $\lambda_i$ is eigenvalue of corresponding density operator. Quantum mutual information constant total correlation of the system, this means that quantum mutual information is the sum of classical correlation $\mathcal{C}(\rho^{AB})$ and quantum correlation $\mathcal{Q}(\rho^{AB})$. So we can get quantum part by\cite{qd_oz}
\begin{equation}
  \mathcal{Q}(\rho^{AB}) = \mathcal{I}(\rho^{AB}) - \mathcal{C}(\rho^{AB}).
\end{equation}

Since, classical correlation is depend on measurement, we used to view part A by measuring part B. Give a set of basis $\{\Pi_k | \Pi_k = I \otimes B_k\}$ to measure B, the state of the system after measuring is
\begin{equation}
  \rho_k = \Pi_k \rho^{AB} \Pi_k / p_k
\end{equation}
here $p_k = \mathrm{tr}_B (\Pi_k \rho^{AB})$ is the probability of measurement\cite{qd_oz}.
Then, the conditional entropy of the final state is
\begin{equation}\label{eq:ce}
  S(\rho^{AB}|\{B_k\}) = \sum_k p_k S(\rho_k),
\end{equation}
Now, the mutual information of this system is\cite{qdx_ali}
\begin{equation}
  \mathcal{I}(\rho^{AB}|\{B_k\}) \equiv S(\rho^A) - S(\rho^{AB}|\{B_k\}).
\end{equation}
Henderson and Vedral\cite{cc_lhvv,cc_vv} gave the explicitly form of classical correlation
\begin{eqnarray}
  \mathcal{C}(\rho^{AB}) &=& \sup_{\{B_k\}} \mathcal{I}(\rho^{AB} | \{B_k\}) \\ \nonumber
  &=& S(\rho^A) - \min_{\{B_k\}} S(\rho^{AB}|\{B_k\}).
\end{eqnarray}

From the above, we can know that the quantum correlation of a bipartite system, which named QD by Olliver and Zurek\cite{qd_oz}, is
\begin{equation}\label{eq:discord}
  \mathcal{Q}(\rho^{AB}) = S(\rho^B) - S(\rho^{AB}) + \min_{\{B_i\}} S(\rho^{AB} | \{B_i\})
\end{equation}

There are a few states that can be solved analytically.
In 2008, Luo\cite{dq_liluo,qdx_luo} got analytical solution of QD for a kind of qubit-qubit system state which is
$\rho = I + \sum_{j=i}^3 c_j \sigma_j \otimes \sigma_j.$
In 2010, Ali with coworkers\cite{qdx_ali} got a more general solution of a kind of state which called X-state, which elements are all zeros except main-diagonal and antidiagonal by similar method.
In 2015, Maldonado-Trapp et.al. gave some supplements to the results of two-qubit X-states QD
\cite{Maldonado-Trapp2015}.
However, it is still a hard task to solve the third term in Eq.(\ref{eq:discord}) for large systems or  general cases.

In order to estimate QD by neural network (NN) well, we should choose an appropriate target function to reduce model complexity, In other word, the simpler taget function is, the better performance we can get. From Eq.(\ref{eq:discord}), we can see that the first two terms are easy to analytically solve, so we just pay attention to the third term of Eq.(\ref{eq:discord}), called optimization term. For convenience, we write it as
\begin{equation}\label{eq:optim}
  c(\rho^{AB}) \equiv \min_{\{B_i\}} S(\rho^{AB} | \{B_i\}).
\end{equation}

\section{Machine Learning Model}
\label{sec:mlm}

Since the original parameters of a two-qubit state are too few to reveal feature directly. Therefore, before feed data to a network, we need a feature transformation as pre-processing to make feature more obvious. We apply a power series transformation by the following rule. Let $n$ is the dimension of the original feature space, and $L$ is the highest power after transforming. So, for a vector $\vec{x} \in \mathbb{R}^n$, the result is $\Phi(\vec{x}) = (1, x_1, x_2, \dots, \phi(\vec{x})_{a_1\cdots a_n}, \dots, x_n^L)^T$, and the element $\phi(\vec{x})_{a_1\dots a_n}$ is
\begin{equation}\label{eq:phib}
  \phi(\vec{x})_{a_1\cdots a_n} = \prod_{i=1}^n x_i^{a_i},\;\; \sum_i a_i \leq L,
\end{equation}
here $a_i$ is the power of the $i$-th base of $\vec{x}$.
And, we list some size of $\Phi$ for some cases in Tab.\ref{tab:fd}.

\begin{table}\center
  \caption{The dimension of the feature space $\Phi$ with highest power $L$, and $n$ is the size of the original parameter space.}
  \label{tab:fd}
\begin{tabular}{cccccccccc}
  \hline\noalign{\smallskip}
  $L$ & 1 & 2 & 3 & 4 & 5 & 6 & 7 & 8 & 9 \\
  \noalign{\smallskip}\hline\noalign{\smallskip}
  $n=7$ & 8 & 36 & 120 & 330 & 792 & 1,716 & 3,432 & 6,435 & 11,440 \\
  $n=9$ & 10 & 55 & 220 & 715 & 2,002 & 5,005 & 11,440 & 24,310 & 48,620 \\
\noalign{\smallskip}\hline
\end{tabular}
\end{table}

\subsection{Estimating by ordinary NN}

\begin{figure}\center
  \includegraphics[scale=0.75]{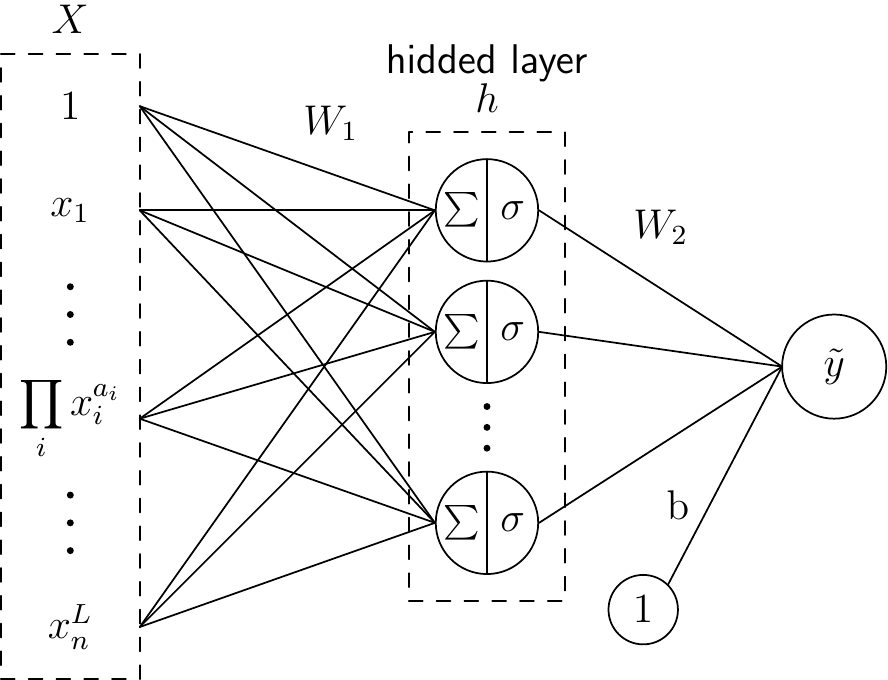}
  \caption{\label{fig:model-nn} An ordinary NN with one hidden layer. $X$ is input data set, $h$ is a hidden layer with $F$ nodes, $\tilde{y}$ is output of this model. $\mathrm{\Sigma}$ means summing the input for each node, and $\sigma$ is sigmoid active function. $W_1$ and $W_2$ are the weight matrices between layers.}
\end{figure}

Firstly, we use an ordinary NN attempt to estimae QD. The NN model we used has only one hidden layer, see Fig.\ref{fig:model-nn}.
Assume we have a raw data set $\mathcal{D} = \{(\vec{x}^{(i)}, c^{(i)}) | i=1,\dots,M \}$, here $M$ is the size of data set.
The input of the NN can be written as a matrix, $X$, forming by feature vectors and $X_{i,j} = \Phi(\vec{x}^{(i)})_j$ is the $i$-th data's $j$-th feature.
The hidden layer $h$ has $F$ nodes.
The values of optimization term, or labels for ML, can be written as a vector $\vec{y}$, here $y_i = c^{(i)}$, is the label of the $i$-th data.
 $W_1$ and $W_2$ are weight matrices of net connections, $\vec{\tilde{y}}$ is the predictive value of the model, and $\sigma$ is the active function , $\sigma(x) = 1/({1+\exp(-x)})$.
The computations of this NN are
\begin{eqnarray*}
  h &=& \sigma(W_1\, X)  \\
  \vec{\tilde{y}} &=& W_2\, h
\end{eqnarray*}

\subsection{NN with Prior Knowledge}

Adding prior knowledge about special problem in the model can help us improve the performance of the model. From the definition of QD, we can see that knowledge about entropy is an important concept in this problem. So, our model should know something about entropy.
From Eq.(\ref{eq:ce}), we can know that $c$ is a sum of von Neumann entropy of a density operator ensemble, $\{p_i, \tilde{\rho}_i\}$, the only problem is that we do not have an effective method to find it. Even though, we can sure that getting $\{p_i, \tilde{\rho}_i\}$ is equivalent getting their eigenvalues, so we have the following form
\begin{equation}
  c = \sum_{i, j} p_i \tilde{\lambda}_{i,j} \log_2 \tilde{\lambda}_{i,j}
\end{equation}
here, $\tilde{\lambda}_{i,j}$ are eigenvalues of density operator $\tilde{\rho}_i$.
thus we replace the active function by a new "active function" 
\begin{equation}\label{eq:eaf}
  E(x) \equiv \left\{ \begin{array}{ll} 0, & x \leq 0; \\ x \log_2 x, & x > 0. \end{array}\right.
\end{equation}
In this NN model, the input of this layer may be the approximate of eigenvalues, ideally. What' more, the change of active function can further reduce the model complexity for previous layer.
\begin{figure}\center
  \includegraphics[scale=0.75]{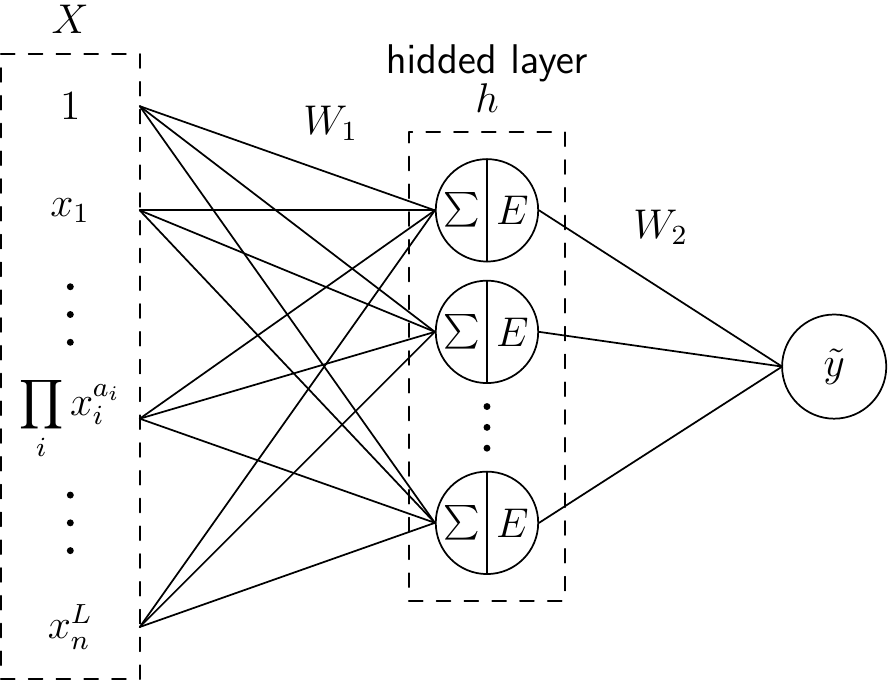}
  \caption{\label{fig:model-pk} A NN model with knowledge about entropy. $X$ is input data set, $h$ is a hidden layer with $F$ nodes, $\tilde{y}$ is output of this model. $\mathrm{\Sigma}$ means summing the input for each node, and $E$ is knoweledge about entropy defined as Eq.(\ref{eq:eaf}). $W_1$ and $W_2$ are the weight matrices between layers.}
\end{figure}

We illustrate this model in Fig.\ref{fig:model-pk}, we call it PKNN for convenience.
The computations of this model are,
\begin{eqnarray*}
  h &=& E(W_1 \, X)  \\
  \tilde{y} &=& W_2 \, h.
\end{eqnarray*}
The second layer realizes the sum in the definition in von Neumann entropy in some way. And, we don't need offset term in this layer according to the definition of entropy.

\subsection{Double Branch NN with Prior Knowledge}

We should note one thing that because of the optimizer, $\min_{\{B_i\}}$, the result of QD is shattered. In other word, there are some, more than one, conditions $g(\vec{x}) \geq 0$, cause the target function $c(\vec{x})$ or eigenvalue $\tilde{\lambda}_{i,j}$ have different analytical forms in different parameter ranges,
\begin{equation}
  c(\vec{x}) = \left\{\begin{array}{ll}
    c_1(\vec{x}), & g_1(\vec{x}) \geq 0; \\
    \vdots       & \vdots \\
    c_n(\vec{x}), & g_n(\vec{x}) > 0.
  \end{array}\right.
\end{equation}
The above can be written in one formula
\begin{equation}
  c(\vec{x}) = \sum_i \theta(g_i(\vec{x}))\; c_i(\vec{x})
\end{equation}
here, $\theta(\cdot)$ is step function
\begin{equation}
  \theta(a) = \left\{\begin{array}{ll}
    0, & a \leq 0; \\
    1, & a > 0.
  \end{array}\right.
\end{equation}
Unfortunately, how much conditions are there in the target function is unknown in advance, therefore, we should choose an appropriate value in an experiment.

\begin{figure}\center
  \includegraphics[scale=0.75]{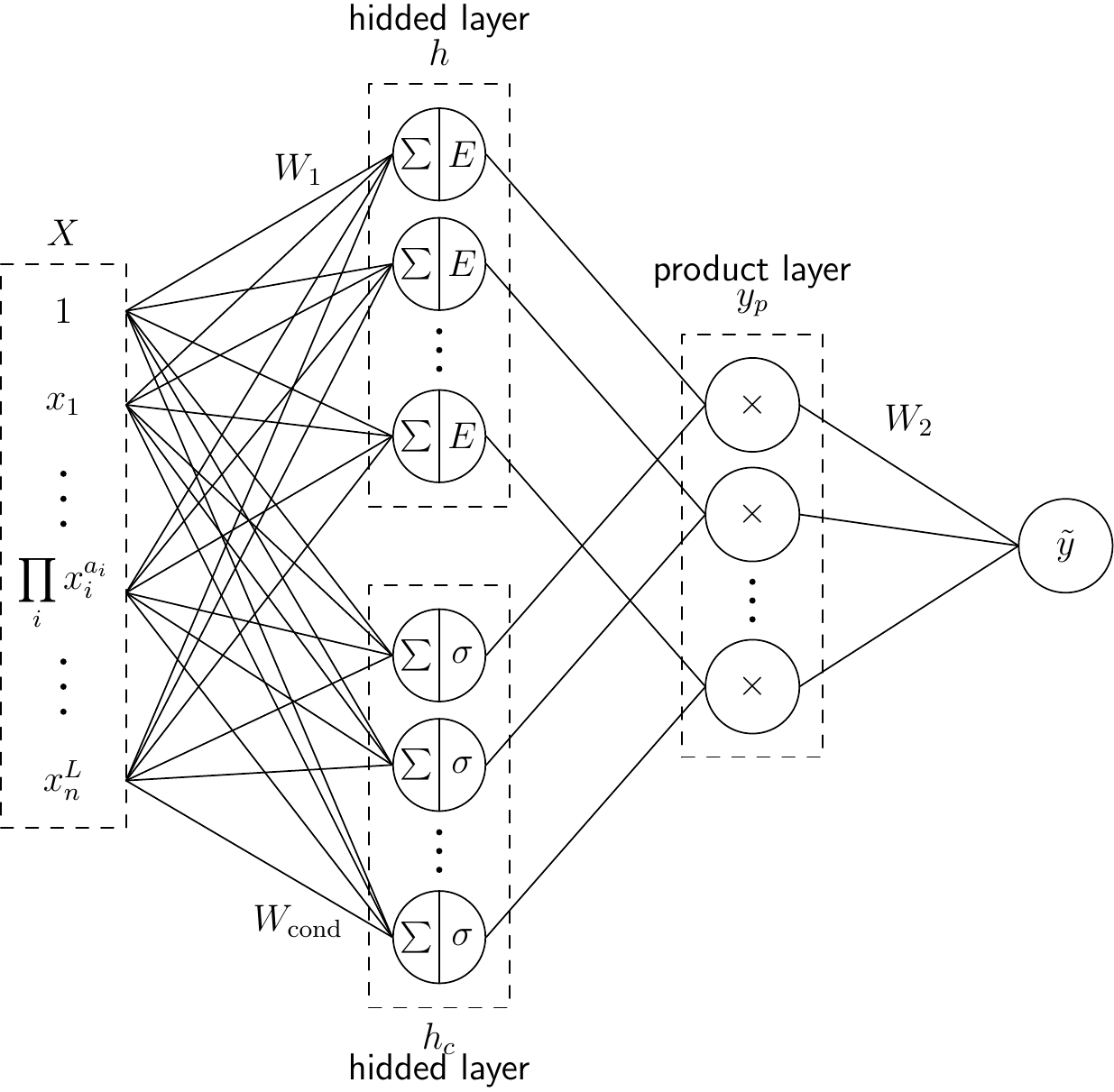}
  \caption{\label{fig:model-db} A double branch NN with conditional function. $X$ is input data set. $h$ is a hidden layer with $F$ nodes for eigenvalues, $h_c$ is a hidden layer with $F$ nodes too for condition function. $\tilde{y}$ is output of this model. $y_p$ is middle layer by multiplying $h$ and $h_c$. $\mathrm{\Sigma}$ meas summation of input, $\sigma$ is sigmoid active function and $E$ is knoweledge about entropy. $W_1$, $W_2$ and $W_\mathrm{cond}$ are the weight matrices between layers.}
\end{figure}

Based on the above discussion, we modified the model again, the new one named double branch neural network (DBNN), see Fig.\ref{fig:model-db}.
In this model, we add condition control factors in a new path. Ideally, the value of a condition function $g$, or output of $h_c$, should belong to $\{0,1\}$. However, step function is hard for training, so we choose sigmoid function, $\sigma$, as alternative. The whole computations of this model are
\begin{eqnarray*}
  h &=& E(W_1 \, X)\\
  h_c &=& \sigma( W_\mathrm{cond} \, X)  \\
  (y_p)_{i,j} &=& h_{i,j} \, (h_c)_{i,j} \\
  \tilde{y} &=& W_2 \, y_p 
\end{eqnarray*}
The layers of $h$ and $h_c$ are used to probe eigenvalues $\tilde{\lambda}_{i,j}$ and condition functions $g(\vec{x})$, then they are an approximation of eigenvalues and conditions.

\section{Result}
\label{sec:res}

We will compare above three models in this section. To evaluate the performance of these models we use mean quadratic loss function as index:
\begin{equation}
  \ell = \frac1M\sum_{i}^M (y^{(i)} - \tilde{y}^{(i)})^2.
\end{equation}
At last, these models are trained by TensorFlow\cite{tensorflow.org}.

\subsection{Result on X-state}
First, we test our models on two-qubit X-states,
\begin{equation}
  \rho^{AB}_X = \left( \begin{array}{cccc}
    \rho_{11} & 0 & 0 & \rho_{14} \\
    0 & \rho_{22} & \rho_{23} & 0 \\
    0 & \rho_{32} & \rho_{33} & 0 \\
    \rho_{41} & 0 & 0 & \rho_{44} \\
  \end{array}\right),
\end{equation}
we can use a 7-dimension real vector, $\vec{x} \in \mathbb{R}^7$, to characterize it. Thus we have $\rho_{11}=x_1,\; \rho_{22}=x_2,\; \rho_{33}=x_3,\; \rho_{44}=1-x_1-x_2-x_3,\; \rho_{14}=x_4 + i\, x_5,\; \rho_{23}=x_6+i\, x_7$. According to the eigenvalues of $\rho^{AB}_X$\cite{qdx_ali}, there are some extra restrictions for those parameters:
\begin{equation}
\begin{array}{l}
  \abs{\rho_{14}}^2 \leq \rho_{11} \rho_{44}, \\
  \abs{\rho_{23}}^2 \leq \rho_{22} \rho_{33}.
\end{array}
\end{equation}
So, for X-state, the optimization term is a function of a 7-D vector:
\[ c(\rho_X) \in \mathcal{H}_X(\vec{x}),\;\;\; \vec{x} \in \mathbb{R}^7\]

In this experiment, we use a training set with $M=6000$ to inhibit over-fitting. These data are numerical solutions come from randomly generated two-qubit X-states.
In order to balance direct feature and the amount of calculation, we choose $L=6$ as the highest power of transformation and the second layer has $F=16$ nodes.
Every model runs 5 times, and runs $3\times 10^5$ steps each time. Then we compare their average performance. $3\times 10^5$ steps can guarantee that loss function will not change obviously in every turns.
Other training parameters are same, those are initial learning rate $\eta = 0.2$ and decaying with 0.98 every 3,000 steps.

\begin{table}\center
  \caption{Results of the 3 models on X-state, last line is average of loss $\overline{\ell}$. BDNN has the best performance, PKNN is second.}
  \label{tab:res}
    \begin{tabular}{cccccc}
    \hline\noalign{\smallskip}
    \multicolumn{2}{c} {NN}  & \multicolumn{2}{c}{PKNN} & \multicolumn{2}{c}{DBNN} \\
    \cline{1-2} \cline{3-4} \cline{5-6} \noalign{\smallskip}
    \shortstack{training\\($\times 10^{-3}$)} & \shortstack{test\\($\times 10^{-3}$)} &
      \shortstack{training\\($\times 10^{-3}$)} & \shortstack{test\\ ($\times 10^{-3}$)} &
      \shortstack{training\\($\times 10^{-3}$)} & \shortstack{test\\($\times 10^{-3}$)} \\
    \noalign{\smallskip}\hline\noalign{\smallskip}
    3.433 & 3.162 & 1.427 & 1.690 & 1.052 & 1.163 \\
    3.435 & 3.167 & 1.071 & 1.157 & 1.069 & 1.266 \\
    3.007 & 2.887 & 1.340 & 1.352 & 1.211 & 1.319 \\
    2.994 & 2.875 & 1.513 & 1.691 & 0.946 & 1.040 \\
    2.973 & 2.839 & 1.475 & 1.443 & 0.991 & 1.114 \\
    \noalign{\smallskip}\hline\noalign{\smallskip}
    3.168 & 2.986 & 1.365 & 1.467 & 1.054 & 1.180 \\
    \noalign{\smallskip}\hline
  \end{tabular}
\end{table}

Tab.\ref{tab:res} shows all results, the last line is the average value of loss function, $\overline{\ell}$. We can see that PKNN is better than an ordinary NN and DBNN has the best performanceon on average. This accords with our expectation.

Random initialization is hard to find best result each time. One of the reasons is that there are lots of local minima in those models. Moreover, the non-monotonic function $E(\cdot)$ makes the the loss function more complex then monotonic activation function. For the same reasons, those models are unstable. Even so, these results can prove our suppose that prior knowledge can improve the performance of NN.

\begin{figure}\center
  \includegraphics[width=0.45\textwidth]{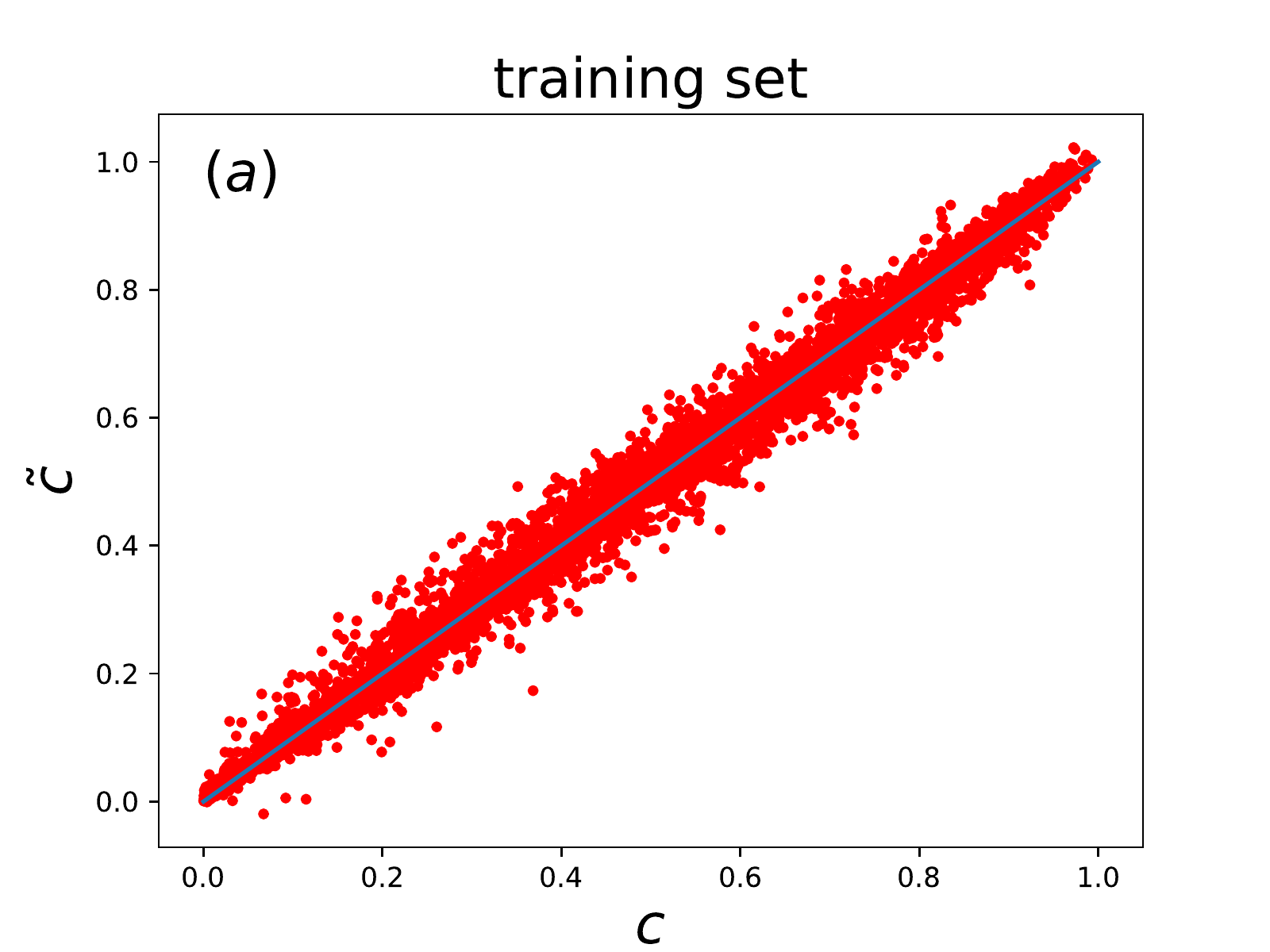}
  \includegraphics[width=0.45\textwidth]{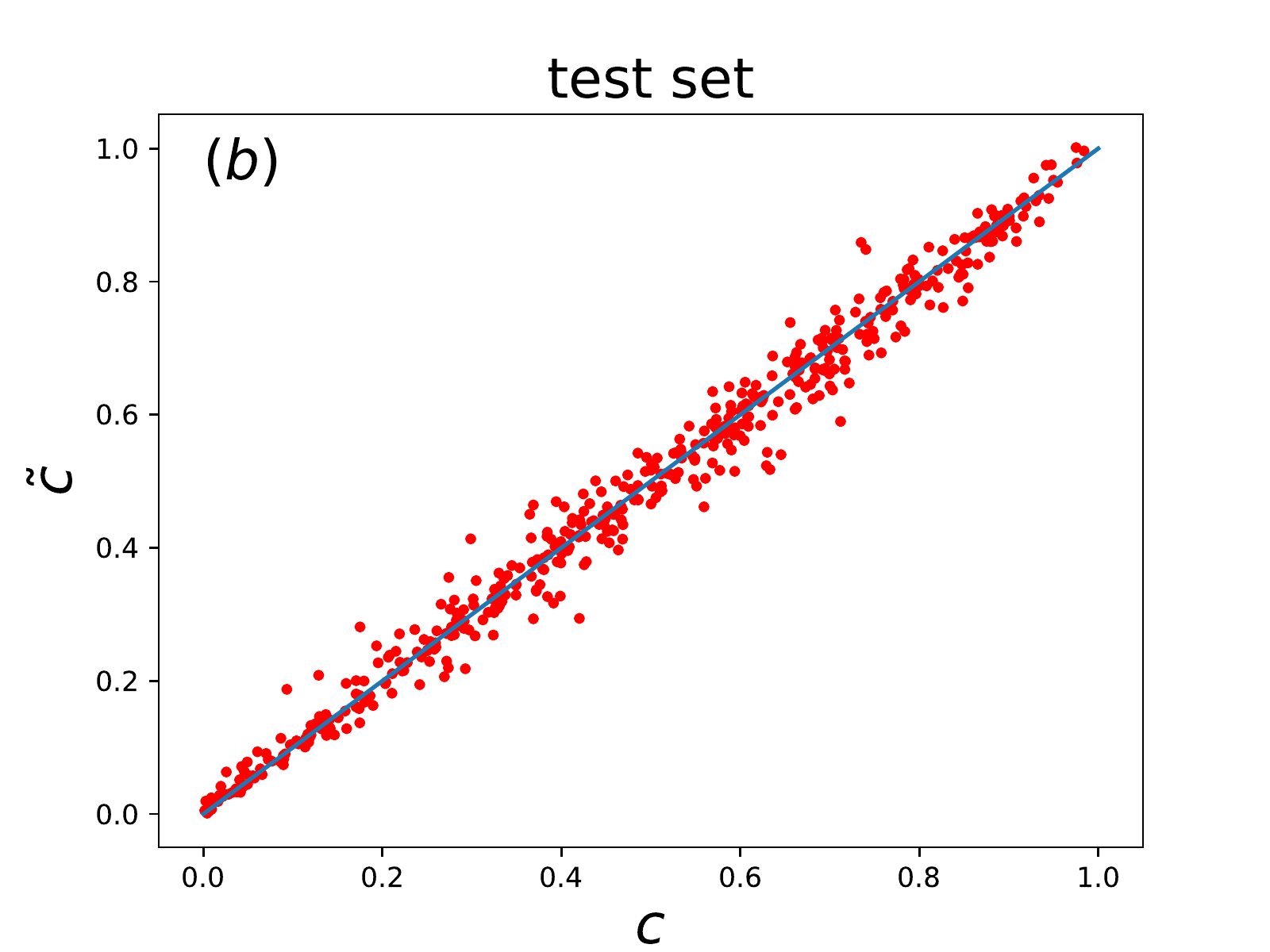}
  \caption{\label{fig:bestResult} The best result of DBNN on X-state. $c$ is theoretical value, $\tilde{c}$ is predictive value. a) the performance on training set, b) the performance on test set. The blue line is $\tilde{c} = c$}
\end{figure}

Then, We show the performance of the best result in Tab.\ref{tab:res}, the result is plotted in Fig.\ref{fig:bestResult}. The horizontal axis is the theoretical value of optimization term or label, $c$, and the vertical axis is the predictive value, $\tilde{c}$ from DBNN.
We can see that the predictive value falls in a bit wide range.

\begin{figure}\center
  \includegraphics[width=0.75\textwidth]{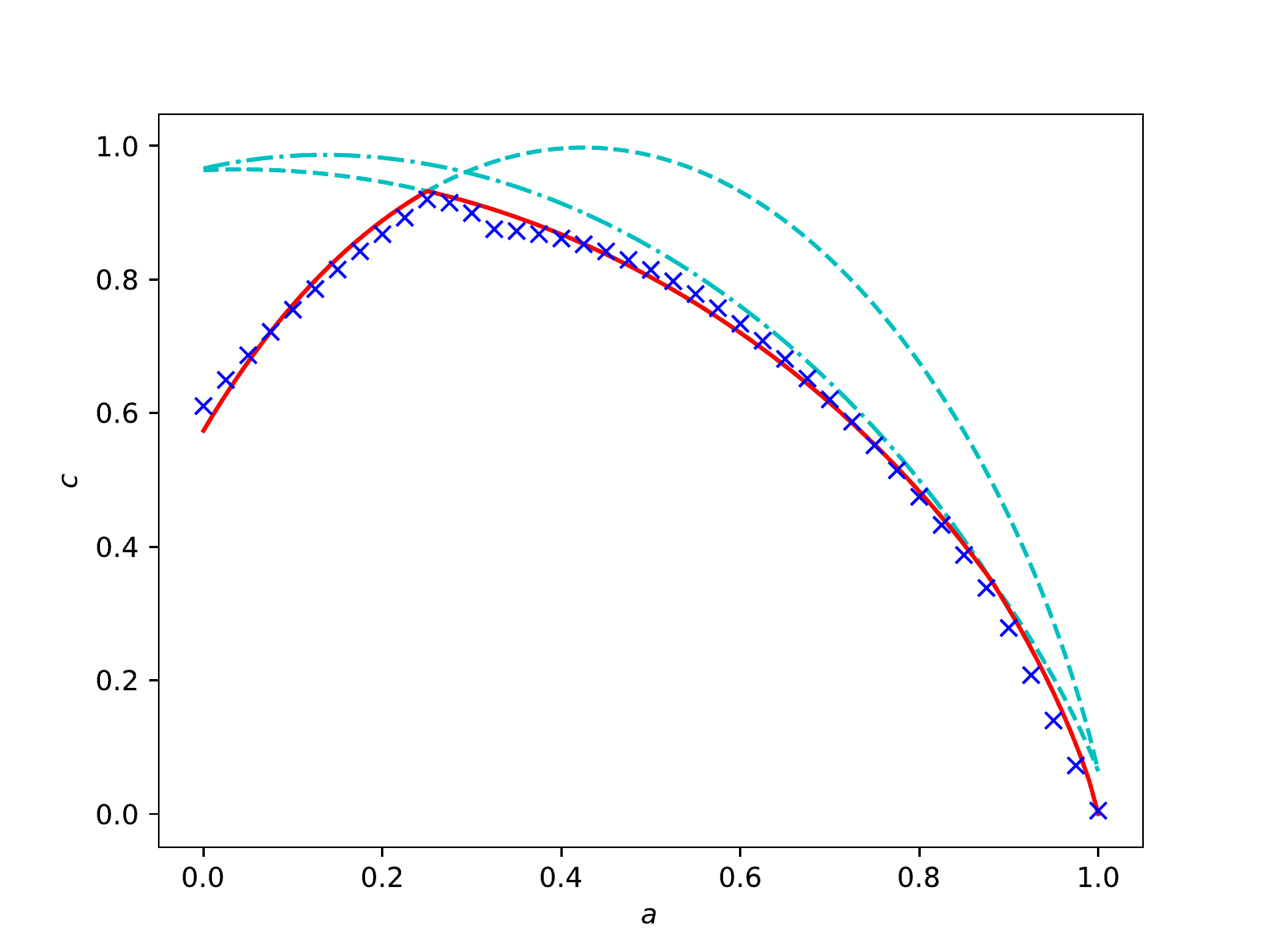}
  \caption{\label{fig:cmpwth} Analytical solution and prediction of a DBNN. The red solid line is Eq.(\ref{eq:as-rho}) and the three cyan dashed lines are $S'(\rho)|_{\theta_1, \theta_2},\, S'(\rho_0)|_{\theta_3}$ and $S'(\rho_0)|_{\theta_4}$. The blue "$\times$" points are our prediction. }
\end{figure}

Next, we test our model on a X-state
\begin{equation}
\rho = \left(\begin{array}{cccc}
0.35-0.35a & 0 & 0 & -0.2+0.2a \\
0 & 0.25+0.25a & -0.15 + 0.6a & 0 \\
0 & -0.15+0.6a & 0.2+0.5a & 0 \\
-0.2+0.2a & 0 & 0 & 0.2-0.2a
\end{array}\right).
\end{equation}
It is easy to get the analytical solution by Ali's result\cite{qdx_ali},
\begin{equation}\label{eq:as-rho}
  c(\rho) = \min_{\{B_i\}} S(\rho|\{B_i\}) = \min \{ S'(\rho)|_{\theta_1, \theta_2},
    S'(\rho_0)|_{\theta_3}, S'(\rho_0)|_{\theta_4} \}
\end{equation}
where
\begin{equation}
  S'(\rho_i)|_{\theta_j} = - \frac{1-\theta_j}{2}\log_2 \frac{1-\theta_j}{2}
     - \frac{1+\theta_j}{2}\log_2 \frac{1+\theta_j}{2}
\end{equation}
\begin{equation}
  S'(\rho)|_{\theta_1, \theta_2} = (0.55+0.15a)S'(\rho_0)|_{\theta_1} 
    + (0.45-0.15a)S'(\rho_1)|_{\theta_2}
\end{equation}
and
\begin{equation}
\begin{split}
  \theta_1 & = \sqrt{\frac{(0.15-0.85a)^2}{(0.55+0.15a)^2}} \\
  \theta_2 & = \sqrt{\frac{(0.05+0.25a)^2}{(0.45-0.15a)^2}} \\
  \theta_3 & = \sqrt{\frac{0.1325-0.62a+0.73a^2}{0.25}} \\
  \theta_3 & = \sqrt{\frac{0.0125-0.02a+0.25a^2}{0.25}}
\end{split}
\end{equation}
We plot the analytical solution and the predictive result of a DBNN in Fig.\ref{fig:cmpwth}. It show that we have a good predictive about QD including condition.

\subsection{Result on real state}

To show the universality of our method, we test it on a special two-qubit state which is researched little, we call it real state. Its elements are all real, $\rho_{ij} \in \mathbb{R}$. We can use 9 independent parameters to characterize it:
\begin{equation}
  \rho^{AB}_R = \left(\begin{array}{cccc}
    \rho_{11} & \rho_{12} & \rho_{13} & \rho_{14} \\
    \rho_{21} & \rho_{22} & \rho_{23} & \rho_{24} \\
    \rho_{31} & \rho_{32} & \rho_{33} & \rho_{34} \\
    \rho_{41} & \rho_{42} & \rho_{43} & \rho_{44}
  \end{array}\right) = 
   \left(\begin{array}{cccc}
    x_1 & x_4 & x_5 & x_6 \\
    x_4 & x_2 & x_7 & x_8 \\
    x_5 & x_7 & x_3 & x_9 \\
    x_6 & x_8 & x_9 & x_0
  \end{array}\right),
  x_0 = 1 - x_1 - x_2 - x_3.
\end{equation}
The solution of this state is unknown, thus, we don't know the range of non-diagonal elements. So we generate this kind of state randomly with following empirical restrictions to improve sampling efficiency,
\begin{equation}
  \abs{\rho_{ij}}^2 \leq \rho_{ii} \rho_{jj}
\end{equation}
and discard the one with negative eigenvalue.
Well, for a real state, the optimization term is a function of a 9-dimensional vector:
\[ c(\rho^{AB}_R) \in \mathcal{H}_R(\vec{x}),\;\;\; \vec{x} \in \mathbb{R}^9\]

The truncated term is still at $L=6$. From previous calculations in Tab.\ref{tab:fd}, we know that the size of first layer is 5,005, it's huge. The size of training set is same as before, $M = 6000$. The initial learning rate $\eta = 0.2$, and decay with 0.96 every 3,000 steps in this experiment. As a transitional model, we don't test PKNN this time.

\begin{table}\center
  \caption{Results of ordinary NN and DBNN on real state, last line is average $\overline{\ell}$. DBNN is better than ordinary NN model.}
  \label{tab:fullReal}
  \begin{tabular}{cccc}
    \hline\noalign{\smallskip}
    \multicolumn{2}{c}{NN}  & \multicolumn{2}{c}{DBNN} \\
    \cline{1-2} \cline{3-4} \noalign{\smallskip}
    \shortstack{training\\($\times 10^{-3}$)} & \shortstack{test\\($\times 10^{-3}$)} &
      \shortstack{training\\($\times 10^{-3}$)} & \shortstack{test\\($\times 10^{-3}$)} \\
	\noalign{\smallskip}\hline\noalign{\smallskip}
    2.200 & 1.996 & 1.104 & 1.183 \\
    2.193 & 1.981 & 1.099 & 1.168 \\
    2.195 & 1.984 & 1.047 & 1.178 \\
    2.196 & 1.992 & 1.162 & 1.237 \\
    2.216 & 2.007 & 1.228 & 1.270 \\
    \noalign{\smallskip}\hline\noalign{\smallskip}
    2.200 & 1.992 & 1.128 & 1.207 \\
    \noalign{\smallskip}\hline
  \end{tabular}
\end{table}

\begin{figure}\center
  \includegraphics[width=0.45\textwidth]{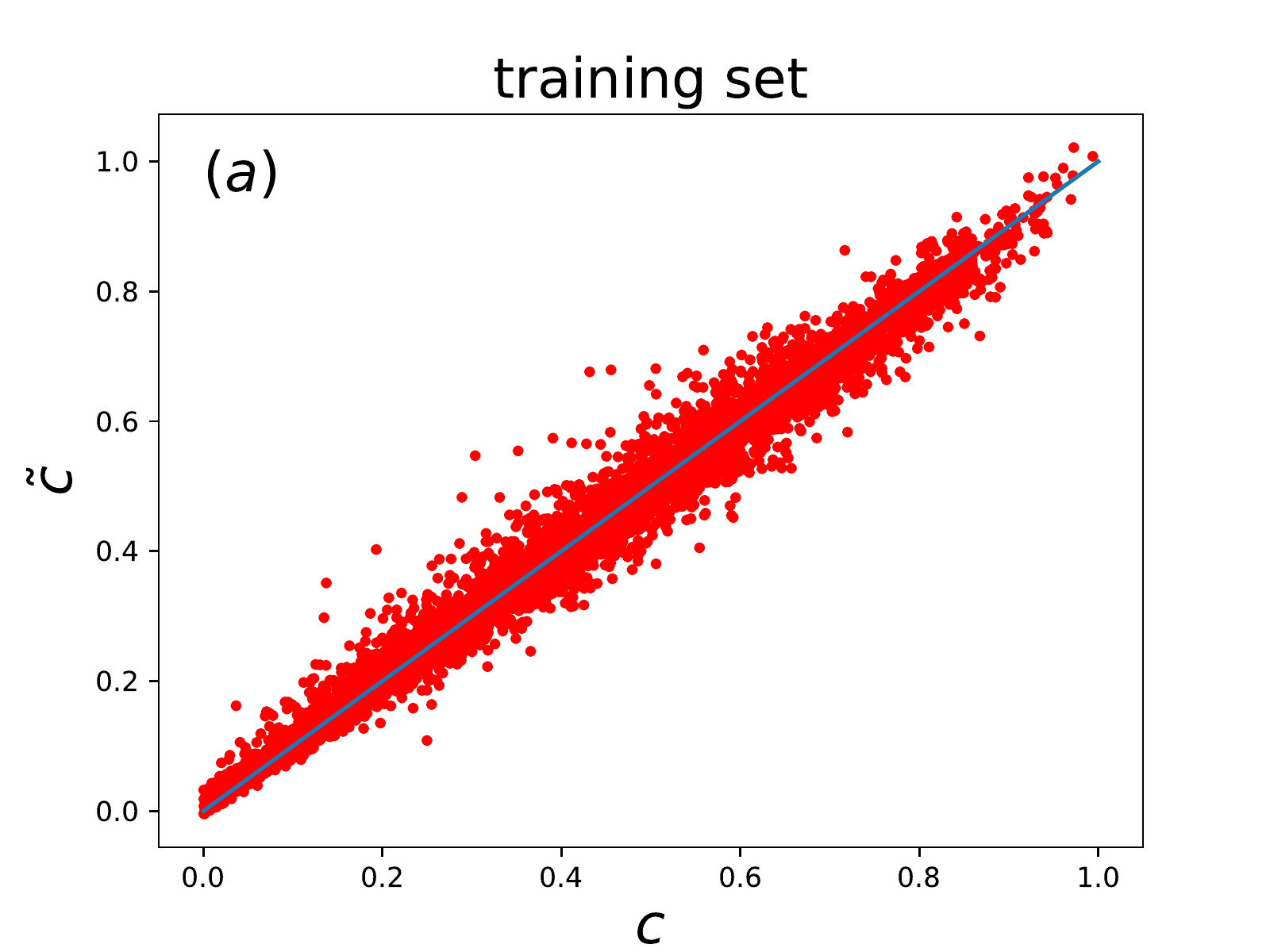}
  \includegraphics[width=0.45\textwidth]{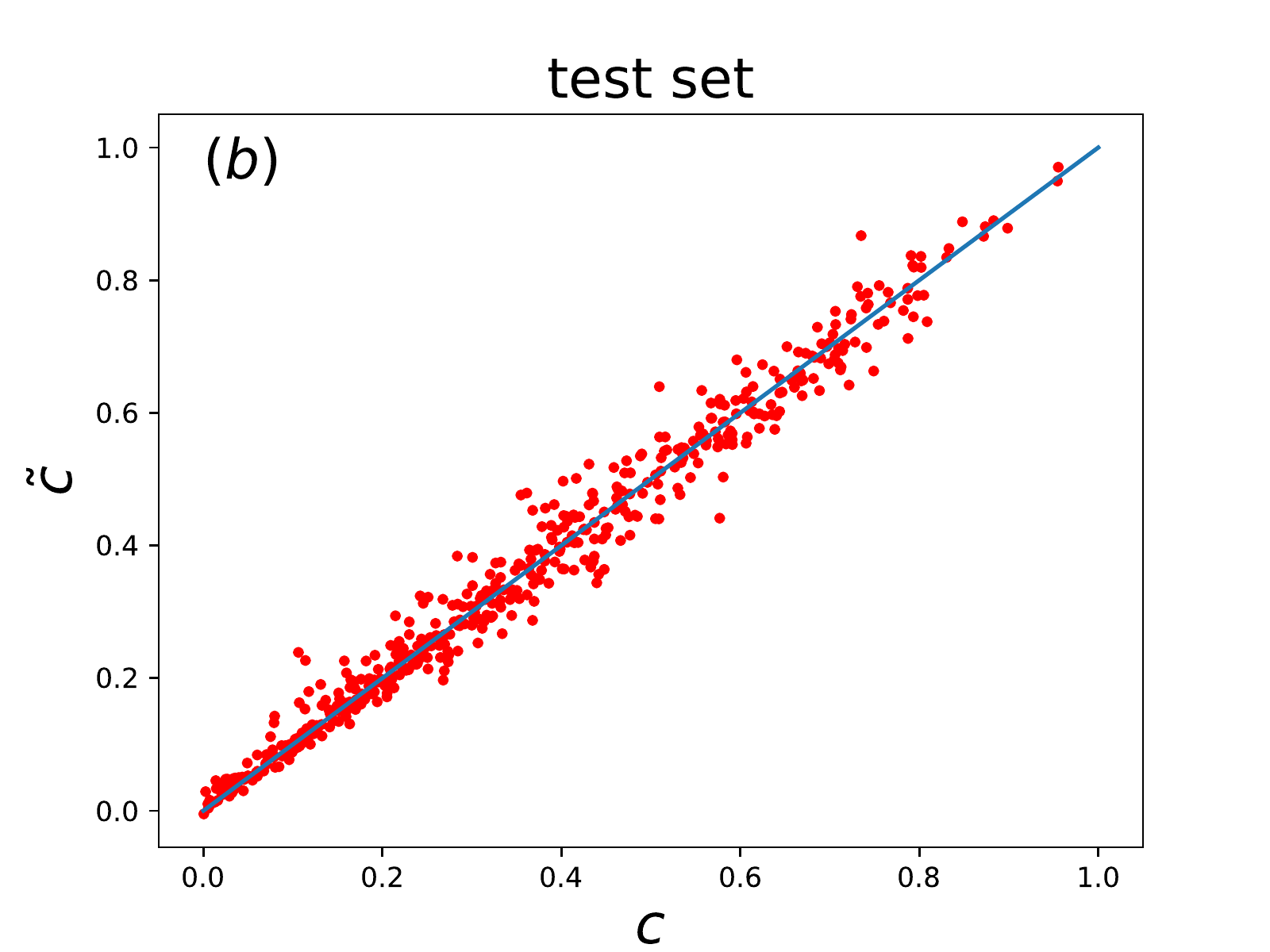}
  \caption{\label{fig:fullReal} The best result of DBNN on real state. $c$ is theoretical value, $\tilde{c}$ is predictive value. a) the performance on training set, b) the performance on test set. The blue line is $\tilde{c} = c$}
\end{figure}

Tab.\ref{tab:fullReal} show all results and Fig.\ref{fig:fullReal} display the best result of DBNN in Tab.\ref{tab:fullReal}. We can see that DBNN has similar loss with X-state on real states training set and test set. The range of data points in Fig.\ref{fig:fullReal} is also similar to Fig.\ref{fig:bestResult}.
By comparing the results of ordinary NN and DBNN, we can see that our modification based on prior knowledge actually improve the performance of network.

\section{Discussion}
\label{sec:diss}

In summary, we build two new models by adding prior knowledge about entropy which are base on the analysis about the problem in this work. Results show that prior knowledge assuredly improve the performance of NN. What's more, our model have an advantages that is the interpretability.

Since the first layer of our model is simple, it will increase exponentially with the number of parameters of a system. What'a more, for a big system, more parameters mean higher power is necessary, then it lead to a huge dimension of the transformed feature space. Thearfore, we hope to find a more effective method to express mathematical expression which should be trainable. If possible, this method maybe extended to the more general physical problems, that, finding the law through "big data" of a phenomenon will provide a new way.

\begin{acknowledgements}
This work is supported by National Natural Science Foundation of China under Grant No. 11375168 and Key Research and Development Plan of Ministry of Science and Technology of China under Grant No. 2018YFB1601402.
\end{acknowledgements}

\bibliography{artref}

\end{document}